\documentclass[12pt]{article}
\usepackage{epsfig}
\begin{document}
%%%%%%%%%%%%%%%%%%%%%%%%%%%%%%%%%%%%%%%%%%%%%%%%%%%%%%%%%%%%%%%%%%%%%%%%%
\title{The possibility to study in-medium modification of $J/\psi$ mesons from their
photoproduction on nuclei near threshold in the case of presence of the LHCb pentaquark states
$P^+_c$ in this photoproduction}
\author{E. Ya. Paryev$^{1,2}$ \\
{\it $^1$Institute for Nuclear Research, Russian Academy of Sciences,}\\
{\it Moscow 117312, Russia}\\
{\it $^2$Institute for Theoretical and Experimental Physics,}\\
{\it Moscow 117218, Russia}}
%==============================================================
%%==============================================================

\renewcommand{\today}{}
\maketitle

\begin{abstract}
   We study the $J/\psi$ photoproduction from nuclei in the near-threshold beam energy region of
   $E_{\gamma}$=5--11 GeV within the nuclear spectral function approach
   by considering incoherent direct (${\gamma}N \to {J/\psi}N$) and two-step
   (${\gamma}p \to P^+_c(4312)$, $P^+_c(4312) \to {J/\psi}p$;
   ${\gamma}p \to P^+_c(4440)$, $P^+_c(4440) \to {J/\psi}p$;
   ${\gamma}p \to P^+_c(4457)$, $P^+_c(4457) \to {J/\psi}p$) $J/\psi$ production processes.
   We calculate the absolute excitation functions for the non-resonant and resonant production of
   $J/\psi$ mesons off $^{12}$C and $^{208}$Pb target nuclei within the different scenarios for in-medium
   modification of the directly photoproduced $J/\psi$ mesons and for branching ratios of the decays
   $P^+_c(4312) \to {J/\psi}p$, $P^+_c(4440) \to {J/\psi}p$ and $P^+_c(4457) \to {J/\psi}p$.
   We show that the non-resonant subthreshold $J/\psi$ production in ${\gamma}A$ reactions reveals
   some sensitivity to employed in-medium modification scenarios
   for $J/\psi$ mesons, which is not masked by the resonantly produced $J/\psi$ mesons,
   only if above branching ratios are less than 5\%.
   We also demonstrate that if these branching ratios more than 5\% then the presence of the $P^+_c(4312)$, $P^+_c(4440)$ and $P^+_c(4457)$ resonances
   in $J/\psi$ photoproduction on nuclei produces above threshold additional enhancements in the
   behavior of the total $J/\psi$ creation cross section on nuclei, which could be also studied in the future
   JLab experiments at the CEBAF facility to provide both further evidence for their existence and valuable
   information on their nature.
\end{abstract}

\newpage

\section*{1. Introduction}

In a recent publication [1] the role of initially claimed [2] by the LHCb Collaboration pentaquark
resonance $P^+_c(4450)$ in $J/\psi$ photoproduction on nuclei at near-threshold incident photon energies
of 5--11 GeV has been investigated within the spectral function approach. The description is based both on
the direct non-resonant (${\gamma}N \to {J/\psi}N$) and on the two-step resonant
(${\gamma}p \to P_c^+(4450)$, $P_c^+(4450) \to {J/\psi}p$) charmonium elementary production mechanisms.
It was found that this role is not essential for the $P^+_c(4450)$ resonance with the spin-parity combination
$J^P=(5/2)^+$ only if the branching fraction $Br[P^+_c(4450) \to {J/\psi}p]$ $\sim$ 1\% and less.
In this case the overall subthreshold $J/\psi$ production in ${\gamma}A$ reactions reveals some sensitivity
to adopted in-medium modification scenarios for the directly photoproduced $J/\psi$ mesons.

  Very recently, the LHCb Collaboration reported the observation of three new narrow pentaquark states [3].
Their updated analysis indicates that previously announced $P^+_c(4450)$ state splits into two narrow peaks,
$P^+_c(4440)$ and $P^+_c(4457)$, in ${J/\psi}p$  invariant mass spectrum of the $\Lambda^0_b \to K^-({J/\psi}p)$
decays. Also a new pentaquark, $P^+_c(4312)$, was discovered in this spectrum. Since the current LHCb
analysis [3] is not sensitive to broad resonances, it can neither confirm nor disprove the existence of the
broad $P^+_c(4380)$ state observed in the original data [2]. The search for the LHCb pentaquark candidates
$P^+_c(4312)$, $P^+_c(4440)$ and $P^+_c(4457)$ through a scan of the cross section of the exclusive reaction
\footnote{$^)$They should appear as structures at $E_{\gamma}=9.44$, 10.04 and 10.12 GeV in this cross section.}$^)$
${\gamma}p \to {J/\psi}p$ from threshold of 8.2 GeV and up to photon energy $E_{\gamma}=11.8$ GeV has been
undertaken recently by the GlueX Collaboration at JLab [4]. No evidence for them has been found with present
statistics and
model-dependent upper limits on branching fractions of $P_c^+(4312) \to {J/\psi}p$, $P_c^+(4440) \to {J/\psi}p$
and $P_c^+(4457) \to {J/\psi}p$ decays were set. However, to get a robust enough information for or against
their existence, improved separate $P_c$ photoproduction measurements at JLab with finer energy binning and
with using the polarization observables are needed (cf. [5, 6] and see also below).

Stimulated by the observation of the hidden-charm pentaquark resonances $P^+_c(4312)$, $P^+_c(4440)$
and $P^+_c(4457)$, we intend to investigate now their role in near-threshold $J/\psi$ photoproduction off nuclei
to shed light on the possibility to study experimentally the modification of the $J/\psi$ meson mass in nuclear medium.
The main purpose of the present paper is to extend the model [1] to $J/\psi$--producing two-step resonant
processes ${\gamma}p \to P_c^+(4312)$, $P_c^+(4312) \to {J/\psi}p$;
${\gamma}p \to P_c^+(4440)$, $P_c^+(4440) \to {J/\psi}p$ and
${\gamma}p \to P_c^+(4457)$, $P_c^+(4457) \to {J/\psi}p$
as well as to incorporate in the calculations new experimental data for the total and differential
cross sections of the ${\gamma}p \to {J/\psi}p$ reaction in the threshold energy region from the GlueX
experiment [4]. Further, we briefly remind the main assumptions of the model [1] and describe, where it is
necessary, the corresponding extensions.
We present also the predictions obtained within this expanded model for the $J/\psi$
excitation functions in ${\gamma}C$ and ${\gamma}Pb$ collisions at near-threshold incident energies.
These predictions can be used for possible extraction of valuable information on the
$J/\psi$ in-medium mass shift from the data which could be taken in a dedicated experiment at the
upgraded up to 12 GeV CEBAF facility.

\section*{2. The framework}

\subsection*{2.1. Direct mechanism of non-resonant $J/\psi$ photoproduction on nuclei}

  At near-threshold photon beam energies below 11 GeV of our interest,
the following direct non-resonant elementary $J/\psi$ production processes
contribute to the $J/\psi$ photoproduction on nuclei [1]:
%formula(1)
\begin{equation}
{\gamma}+p \to J/\psi+p,
\end{equation}
%formula(2)
\begin{equation}
{\gamma}+n \to J/\psi+n.
\end{equation}
In line with [1], we approximate the in-medium local mass $m^*_{{J/\psi}}({\bf r})$ of the $J/\psi$ mesons,
participating in the production processes (1), (2), with their average in-medium mass
$<m^*_{{J/\psi}}>$ defined as:
%formula(3)
\begin{equation}
<m^*_{{J/\psi}}>=m_{{J/\psi}}+V_0\frac{<{\rho_N}>}{{\rho_0}}.
\end{equation}
Here, $m_{{J/\psi}}$ is the free space rest mass of a ${J/\psi}$ meson, $V_0$ is its effective
scalar potential (or its in-medium mass shift) at normal nuclear matter density ${\rho_0}$,
$<{\rho_N}>$ is the average nucleon density. For target nuclei $^{12}$C
and $^{208}$Pb, considered in [1] and here, the ratio $<{\rho_N}>/{\rho_0}$
is approximately equal to 0.5 and 0.8, respectively. In what follows, in line with [1]
for the $J/\psi$ mass shift at normal nuclear matter density $V_0$ we will adopt the following
options: i) $V_0=0$, ii) $V_0=-25$ MeV, iii) $V_0=-50$ MeV, iv) $V_0=-100$ MeV, and v) $V_0=-150$ MeV
as well as will neglect the modification of the outgoing nucleon mass in nuclear matter.

Then, the total cross section for the production of ${J/\psi}$ mesons
on nuclei in the direct non-resonant processes (1) and (2) can be represented as follows [1]:
%formula(4)
\begin{equation}
\sigma_{{\gamma}A\to {J/\psi}X}^{({\rm dir})}(E_{\gamma})=I_{V}[A,\sigma_{{J/\psi}N}]
\left<\sigma_{{\gamma}N \to {J/\psi}N}(E_{\gamma})\right>_A,
\end{equation}
where the effective number of target nucleons participating in the primary ${\gamma}N \to {J/\psi}N$
processes, $I_{V}[A,\sigma_{{J/\psi}N}]$, and
"in-medium" total cross section for the production of ${J/\psi}$
with reduced mass $<m^*_{{J/\psi}}>$ in reactions (1) and (2)
$\sigma_{{\gamma}N\to {J/\psi}N}(\sqrt{s},<m^*_{{J/\psi}}>)$
at the ${\gamma}N$ center-of-mass energy $\sqrt{s}$,
averaged over target nucleon binding and Fermi motion,
$\left<\sigma_{{\gamma}N \to {J/\psi}N}(E_{\gamma})\right>_A$,
are defined by Eqs. (5) and (6) in [1], respectively
\footnote{$^)$In equation (4) it is assumed, as previously in [1], that the $J/\psi$ meson production
cross sections in ${\gamma}p$ and ${\gamma}n$ interactions are the same.}$^)$
.
As in Ref. [1], for the $J/\psi$--nucleon absorption cross section $\sigma_{{J/\psi}N}$ we will use
in our calculations the value $\sigma_{{J/\psi}N}=3.5$ mb.

  Also, as previously in [1], we suggest that the "in-medium" cross section
$\sigma_{{\gamma}N \to {J/\psi}N}({\sqrt{s}},<m^*_{J/\psi}>)$ for $J/\psi$ production in processes (1) and (2)
is equivalent to the vacuum cross section $\sigma_{{\gamma}N \to {J/\psi}N}({\sqrt{s}},m_{J/\psi})$ in which
the free $J/\psi$ mass $m_{J/\psi}$ is replaced by its average in-medium mass $<m^*_{{J/\psi}}>$ as defined by
equation (3) and the free space center-of-mass energy squared s, presented by the formula (9) below,
is replaced by the in-medium expression
%formula(5)
\begin{equation}
  s=(E_{\gamma}+E_t)^2-({\bf p}_{\gamma}+{\bf p}_t)^2.
\end{equation}
Here, $E_{\gamma}$ and ${\bf p}_{\gamma}$ are the energy and momentum of the incident photon,
$E_t$ and ${\bf p}_{t}$ are the total energy and
momentum of the struck target nucleons involved in the elementary processes (1) and (2).
The quantity $E_t$ is determined by Eq. (8) in Ref. [1].
 For the free total cross section $\sigma_{{\gamma}N \to {J/\psi}N}({\sqrt{s}},m_{J/\psi})$
at photon energies $E_{\gamma} \le $ 22 GeV we have used the following expression,
based on the near-threshold predictions of the two gluon and three gluon exchange model [7]:
%formula(6)
\begin{equation}
\sigma_{{\gamma}N \to {J/\psi}N}({\sqrt{s}},m_{J/\psi})= \sigma_{2g}({\sqrt{s}},m_{J/\psi})+
\sigma_{3g}({\sqrt{s}},m_{J/\psi}),
\end{equation}
where
%formula(7)
\begin{equation}
\sigma_{2g}({\sqrt{s}},m_{J/\psi})=a_{2g}(1-x)^2\left[\frac{{\rm e}^{bt^+}-{\rm e}^{bt^-}}{b}\right],
\end{equation}
%formula(8)
\begin{equation}
\sigma_{3g}({\sqrt{s}},m_{J/\psi})=a_{3g}(1-x)^0\left[\frac{{\rm e}^{bt^+}-{\rm e}^{bt^-}}{b}\right]
\end{equation}
and
%formula(9)
\begin{equation}
  x=(s_{\rm thr}-m^2_N)/(s-m^2_N); \,\,\,\,
  s_{\rm thr}=(m_{J/\psi}+m_N)^2, \,\,\,\,
  s=(E_{\gamma}+m_N)^2-{\bf p}_{\gamma}^2.
\end{equation}
Here, $t^+$ and $t^-$ are the maximal and minimal values of the squared four-momentum transfer $t$
between the incident photon and the outgoing $J/\psi$ meson. They correspond to the $t$ where the
$J/\psi$ is produced at angles of 0$^{\circ}$ and 180$^{\circ}$ in ${\gamma}p$ c.m.s., respectively,
and can be expressed through the energies and momenta of the initial photon and the $J/\psi$ meson,
$E^*_{\gamma}, p^*_{\gamma}$ and $E^*_{J/\psi}, p^*_{J/\psi}$, in this system in the following way:
%formula(10)
\begin{equation}
t^{\pm}=m_{J/\psi}^2-2E^*_{\gamma}(m_N^2)E^*_{J/\psi}(m_{J/\psi}){\pm}2p^*_{\gamma}(m_N^2)p^*_{J/\psi}(m_{J/\psi}),
\end{equation}
where
%FORMULA(11)
\begin{equation}
p_{\gamma}^*(m_{N}^2)=\frac{1}{2\sqrt{s}}\lambda(s,0,m_{N}^2),
\end{equation}
%FORMULA(12)
\begin{equation}
p^*_{J/\psi}(m_{J/\psi})=\frac{1}{2\sqrt{s}}\lambda(s,m_{J/\psi}^{2},m_N^2)
\end{equation}
and
%formula(13)
\begin{equation}
E^*_{\gamma}(m_N^2)=p^*_{\gamma}(m_N^2), \,\,\,\,
E^*_{J/\psi}(m_{J/\psi})=\sqrt{m^2_{J/\psi}+[p^*_{J/\psi}(m_{J/\psi})]^2};
\end{equation}
%FORMULA (14)
\begin{equation}
\lambda(x,y,z)=\sqrt{{\left[x-({\sqrt{y}}+{\sqrt{z}})^2\right]}{\left[x-
({\sqrt{y}}-{\sqrt{z}})^2\right]}}.
\end{equation}
%%%%%%%%%%%%%%%%%%%%%%%%%%%%%%%%%%%%%%%%%%%%%%%%%%%%%%%%%%%
\begin{figure}[htb]
\begin{center}
\includegraphics[width=16.0cm]{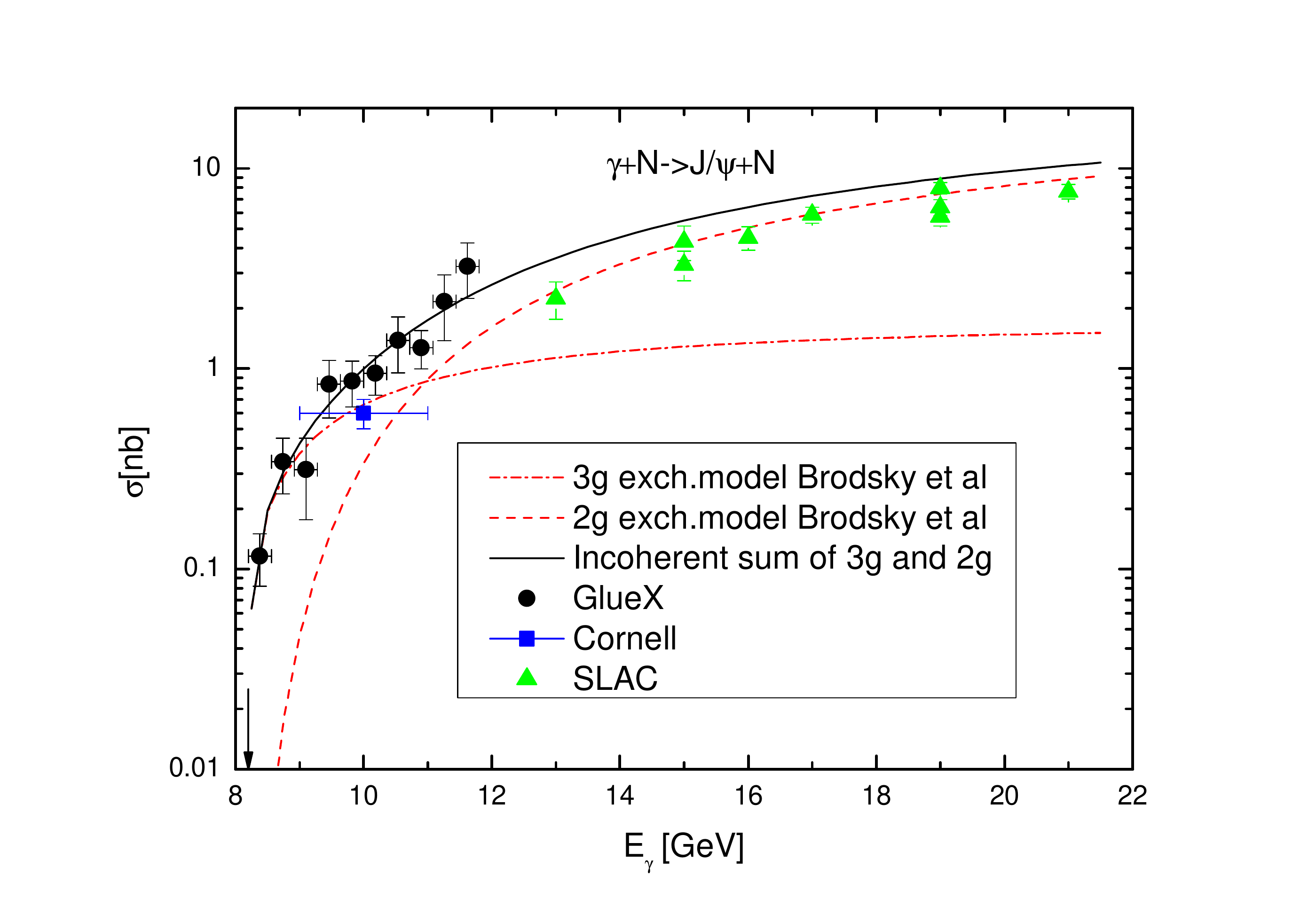}
\vspace*{-2mm} \caption{(color online) The non-resonant total cross section for the reaction
${\gamma}N \to {J/\psi}N$ as function of photon energy. Dashed
and dotted-dashed curves: respectively, calculations on the basis of the two gluon
and three gluon exchange model [7].
Solid curve: incoherent sum of the above two calculations.
The GlueX and SLAC experimental data are from Ref. [4]. The Cornell data point is from Ref. [8].
The arrow indicates the threshold
energy for direct $J/\psi$ photoproduction on a free target nucleon being at rest.}
\label{void}
\end{center}
\end{figure}
%%%%%%%%%%%%%%%%%%%%%%%%%%%%%%%%%%%%%%%%%%%%%%%%%%%%%%%%%%%%%%%%%%%%%%%%%%%%%%%%%%%%%%%%%%%%%
The elementary cross section (6)--(8) was used in the calculations of the non-resonant $J/\psi$
photoproduction off nuclei presented below. In this case the reaction ${\gamma}N \to {J/\psi}N$
goes in medium on an off-shell target nucleon. Then, according to above-mentioned, instead of
the quantities $m_N^2$ and $m_{J/\psi}$, appearing in Eqs. (11) and (9), (12), (13),
we should use the difference $E_t^2-p_t^2$ and $J/\psi$ average in-medium mass $<m^*_{{J/\psi}}>$,
respectively. And instead of the quantity $s$, appearing in Eqs. (9), (11), (12), one needs to adopt its
in-medium expression (5).
Parameter $b$ in Eqs. (7), (8) is an exponential $t$-slope of the differential cross
section of the reaction ${\gamma}p \to {J/\psi}p$ in the near-threshold energy region [7].
According to [4], $b\approx$1.67 GeV$^{-2}$. We will
use this value in our calculations. The normalization coefficients $a_{2g}$ and $a_{3g}$ are determined
assuming that incoherent sum (6) of both channels saturates the total experimental cross section of the
reaction ${\gamma}p \to {J/\psi}p$ measured at GlueX [4] at photon energies around 8.38 and 11.62 GeV,
at which the contribution of the pentaquark resonances $P^+_c(4312)$, $P^+_c(4440)$ and $P^+_c(4457)$
in $J/\psi$ photoproduction on the proton is insignificant (see Fig. 2 given below). As a result, we get
that $a_{2g}=44.780$ nb/GeV$^2$ and $a_{3g}=2.816$ nb/GeV$^2$. Fig. 1 shows that only the combination of the
two gluon and three gluon exchange cross sections fits well the GlueX near-threshold data [4].

\subsection*{2.2. Two-step mechanism of resonant $J/\psi$ photoproduction on nuclei}

  At photon energies below 11 GeV, an incident photon can produce a $P^+_c(4312)$, $P^+_c(4440)$
and $P^+_c(4457)$ resonances in the first inelastic collision with an intranuclear proton
\footnote{$^)$Let us remind that the threshold (resonant) energies $E^{\rm R1}_{\gamma}$,
$E^{\rm R2}_{\gamma}$ and $E^{\rm R3}_{\gamma}$ for the photoproduction of these resonances
with pole masses $M_{c1}=4311.9$ MeV, $M_{c2}=4440.3$ MeV and $M_{c3}=4457.3$ MeV
on a free target proton being at rest are $E^{\rm R1}_{\gamma}=9.44$ GeV,
$E^{\rm R2}_{\gamma}=10.04$ GeV and $E^{\rm R3}_{\gamma}=10.12$ GeV, respectively.}$^)$
:
%formula(16)
\begin{eqnarray}
{\gamma}+p \to P^+_c(4312),\nonumber\\
{\gamma}+p \to P^+_c(4440),\nonumber\\
{\gamma}+p \to P^+_c(4457).
\end{eqnarray}
Then the produced intermediate pentaquark resonances can decay into the $J/\psi$ and $p$:
%formula(17)
\begin{eqnarray}
P^+_c(4312) \to J/\psi+p,\nonumber\\
P^+_c(4440) \to J/\psi+p,\nonumber\\
P^+_c(4457) \to J/\psi+p.
\end{eqnarray}
The branching ratios $Br[P^+_{ci} \to {J/\psi}p]$
\footnote{$^)$Here, $i=$1, 2, 3 and $P^+_{c1}$, $P^+_{c2}$, $P^+_{c3}$ stand for $P^+_c(4312)$, $P^+_c(4440)$,
$P^+_c(4457)$, respectively.}$^)$
of these decays have not been determined yet.
Model-dependent upper limits on these ratios of 4.6\%, 2.3\% and 3.8\% for $P^+_c(4312)$, $P^+_c(4440)$ and
$P^+_c(4457)$, assuming for each $P^+_{ci}$ spin-parity combination $J^P=(3/2)^-$, were set by the GlueX
Collaboration [4]. They become a factor of 5 smaller if $J^P=(5/2)^+$ is supposed. Therefore, following Refs. [1, 4]
as well as Refs. [5, 6, 9, 10], we will employ in our study for these three ratios three following main options:
$Br[P^+_{ci} \to {J/\psi}p]=1$, 2 and 3\%.
To get a better impression of the size of the effect of branching
fractions $Br[P^+_{ci} \to {J/\psi}p]$ on the resonant $J/\psi$ yield in ${\gamma}$C $\to {J/\psi}X$ and
${\gamma}$Pb $\to {J/\psi}X$ reactions, we will also calculate this yield additionally assuming that
all these three branching fractions are equal to 5\% [1] and 10\% [5, 11].

According to [1],
most of the $P^+_c(4312)$, $P^+_c(4440)$ and $P^+_c(4457)$ resonances, having vacuum total decay widths
in their rest frames $\Gamma_{c1}=9.8$ MeV, $\Gamma_{c2}=20.6$ MeV and $\Gamma_{c3}=6.4$ MeV [3], respectively,
decay to $J/\psi$ and $p$ outside the target nuclei considered.
As in [1], their free spectral functions are described by the non-relativistic Breit-Wigner distributions:
%formula(17)
\begin{equation}
S_{ci}(\sqrt{s},\Gamma_{ci})=\frac{1}{2\pi}\frac{\Gamma_{ci}}{(\sqrt{s}-M_{ci})^2+{\Gamma}_{ci}^2/4},\,\,\,
i=1, 2, 3;
\end{equation}
where $\sqrt{s}$ is the total ${\gamma}p$ center-of-mass energy given by Eq.~(9)
\footnote{$^)$When calculating the excitation functions for resonant production of
$P^+_c(4312)$, $P^+_c(4440)$ and $P^+_c(4457)$ states in reactions (15)
on $^{12}$C and $^{208}$Pb targets in the "free" $P^+_c(4312)$, $P^+_c(4440)$ and $P^+_c(4457)$
spectral function scenario (see Fig.~4 below), this energy is determined in line with Eq.~(5).}$^)$
and $S_{c1}$,  $S_{c2}$, $S_{c3}$ correspond to the $P^+_c(4312)$, $P^+_c(4440)$ and $P^+_c(4457)$, respectively.
Following [1], we assume that the in-medium $P^+_c(4312)$, $P^+_c(4440)$ and $P^+_c(4457)$
resonances spectral functions $S_{c1}(\sqrt{s},\Gamma_{\rm med}^{c1})$,
$S_{c2}(\sqrt{s},\Gamma_{\rm med}^{c2})$ and $S_{c3}(\sqrt{s},\Gamma_{\rm med}^{c3})$
are also described by the Breit-Wigner formula (17), respectively, with a total in-medium
widths $\Gamma_{\rm med}^{c1}$, $\Gamma_{\rm med}^{c2}$ and $\Gamma_{\rm med}^{c3}$
in their rest frames, obtained as a sum of the vacuum decay widths, $\Gamma_{c1}$, $\Gamma_{c2}$,
$\Gamma_{c3}$, and an
in-medium contributions due to $P^+_{ci}N$ inelastic collisions -- averaged over local nucleon density
$\rho_N({\bf r})$ collisional widths $<\Gamma_{{\rm coll},c1}>$, $<\Gamma_{{\rm coll},c2}>$ and
$<\Gamma_{{\rm coll},c3}>$:
%formula(18)
\begin{equation}
\Gamma_{\rm med}^{ci}=\Gamma_{ci}+<\Gamma_{{\rm coll},ci}>, \,\,\,i=1, 2, 3;
\end{equation}
where, according to [1], the average collisional width $<\Gamma_{{\rm coll},ci}>$ reads:
%formula(19)
\begin{equation}
<\Gamma_{{\rm coll},ci}>={\gamma_c}{v_c}{\sigma_{P_{ci}N}}<\rho_N>.
\end{equation}
Here, $\sigma_{P_{ci}N}$ is the $P^+_{ci}$--nucleon inelastic cross section
\footnote{$^)$We use in the following calculations $\sigma_{P_{ci}N}=33.5$ mb [1] for each $i=$1, 2, 3.}$^)$
and the Lorentz $\gamma$-factor
$\gamma_c$ and the velocity $v_c$ of the resonance $P_{ci}^+$ in the nuclear rest frame are given by
formula (16) in [1].

Exploring the molecular scenario of $P^+_c(4312)$, $P^+_c(4440)$ and $P^+_c(4457)$ states
\footnote{$^)$In this scenario, due to the closeness of the observed
$P^+_c(4312)$ and $P^+_c(4440)$, $P^+_c(4457)$ masses to the ${\Sigma^+_c}{\bar D}^0$ and
${\Sigma^+_c}{\bar D}^{*0}$ thresholds of 4317.7 and 4459.9 MeV [12], respectively, the $P^+_c(4312)$
resonance can be, in particular, considered as S-wave ${\Sigma^+_c}{\bar D}^0$ bound state, while the
$P^+_c(4440)$ and $P^+_c(4457)$ as S-wave ${\Sigma^+_c}{\bar D}^{*0}$ bound molecular states
[5, 10, 13--24].}$^)$
,
in which their spins-parities are $J^P=(1/2)^-$ for $P^+_c(4312)$, $J^P=(1/2)^-$ for $P^+_c(4440)$ and
$J^P=(3/2)^-$ for $P^+_c(4457)$ [5, 10, 13, 14], we can describe
the free Breit-Wigner total cross sections for their production
in reactions (15) on the basis of the spectral functions (17), provided that the branching
ratios $Br[P^+_{ci} \to {\gamma}p]$ are known, as follows [9, 11]:
%formula(20)
\begin{equation}
\sigma_{{\gamma}p \to P^+_{ci}}(\sqrt{s},\Gamma_{ci})=f_{ci}\left(\frac{\pi}{p^*_{\gamma}}\right)^2
Br[P^+_{ci} \to {\gamma}p]S_{ci}(\sqrt{s},\Gamma_{ci})\Gamma_{ci},\,\,\,i=1, 2, 3;
\end{equation}
where the center-of-mass momentum in the initial ${\gamma}p$ channel, $p^*_{\gamma}$,
is defined above by Eq. (11) and the ratios of spin factors $f_{c1}=1$, $f_{c2}=1$, $f_{c3}=2$.

Following [6, 11, 25], we assume that the $P^+_c(4312)$ $(1/2)^-$, $P^+_c(4440)$ $(1/2)^-$  and
$P^+_c(4457)$ $(3/2)^-$ decays to ${J/\psi}p$ are dominated by the lowest
partial waves with relative orbital angular momentum $L=0$. In this case the branching
ratios $Br[P^+_c(4312) \to {\gamma}p]$, $Br[P^+_c(4440) \to {\gamma}p]$ and $Br[P^+_c(4457) \to {\gamma}p]$
can be expressed using the vector-meson dominance model,
respectively, via the branching fractions $Br[P^+_c(4312) \to {J/\psi}p]$,
$Br[P^+_c(4440) \to {J/\psi}p]$ and $Br[P^+_c(4457) \to {J/\psi}p]$ as follows [6, 9, 11, 25]:
%formula(21)
\begin{equation}
Br[P^+_{ci} \to {\gamma}p]=4{\pi}{\alpha}\left(\frac{f_{J/\psi}}{m_{J/\psi}}\right)^2f_{0,ci}
\left(\frac{p^*_{\gamma,ci}}{p^*_{J/\psi,ci}}\right)
Br[P^+_{ci} \to {J/\psi}p],\,\,\,i=1, 2, 3;
\end{equation}
where $f_{J/\psi}=$ 280 MeV is the $J/\psi$ decay constant, $\alpha=$1/137 is the electromagnetic
fine structure constant and
%FORMULA(22)
\begin{equation}
p_{\gamma,ci}^*=\frac{1}{2M_{ci}}\lambda(M_{ci}^2,0,m_{N}^2),\,\,\,
p^*_{J/\psi,ci}=\frac{1}{2M_{ci}}\lambda(M_{ci}^2,m_{J/\psi}^{2},m_N^2),\,\,\,
\end{equation}
%FORMULA(23)
\begin{equation}
f_{0,ci}=\frac{2}{2+{\gamma}^2_{ci}},\,\,\,\,\,{\gamma}^2_{ci}=1+p^{*2}_{J/\psi,ci}/m^2_{J/\psi}.
\end{equation}
Accounting for that ($p_{\gamma,c1}^*,p^*_{J/\psi,c1},f_{0,c1})=$(2.054~GeV/c, 0.658~GeV/c, 0.657),
($p_{\gamma,c2}^*,p^*_{J/\psi,c2},f_{0,c2})=\\$(2.121~GeV/c, 0.810~GeV/c, 0.652) and
($p_{\gamma,c3}^*,p^*_{J/\psi,c3},f_{0,c3})=$(2.130~GeV/c, 0.828~GeV/c, 0.651), we get from Eq. (21):
%formula(24)
\begin{eqnarray}
Br[P^+_c(4312) \to {\gamma}p]=1.52\cdot10^{-3}Br[P^+_c(4312) \to {J/\psi}p],\nonumber\\
Br[P^+_c(4440) \to {\gamma}p]=1.26\cdot10^{-3}Br[P^+_c(4440) \to {J/\psi}p],\nonumber\\
Br[P^+_c(4457) \to {\gamma}p]=1.24\cdot10^{-3}Br[P^+_c(4457) \to {J/\psi}p].
\end{eqnarray}
The free resonant total cross sections
$\sigma_{{\gamma}p \to P^+_{ci}\to {J/\psi}p}(\sqrt{s},\Gamma_{ci})$ for $J/\psi$ production in the
two-step processes (15), (16) can be represented as follows [1]:
%FORMULA (25)
\begin{equation}
\sigma_{{\gamma}p \to P^+_{ci}\to {J/\psi}p}(\sqrt{s},\Gamma_{ci})=
\sigma_{{\gamma}p \to P^+_{ci}}(\sqrt{s},\Gamma_{ci})\theta[\sqrt{s}-(m_{J/\psi}+m_N)]
Br[P^+_{ci} \to {J/\psi}p],\,\,\,i=1, 2, 3.
\end{equation}
According to Eqs.~(20) and (21) these cross sections are proportional to $Br^2[P^+_{ci} \to {J/\psi}p]$.

In line with [1], we get the following expression for the $J/\psi$ total cross section for
${\gamma}A$ reactions from the production/decay chains (15), (16):
%formula(26)
\begin{equation}
\sigma_{{\gamma}A\to {J/\psi}X}^{({\rm sec})}(E_{\gamma})=
\sum_{i=1}^3\sigma_{{\gamma}A\to P^+_{ci}\to{J/\psi}p}^{({\rm sec})}(E_{\gamma}),
\end{equation}
where
%formula(27)
\begin{equation}
\sigma_{{\gamma}A\to P^+_{ci}\to{J/\psi}p}^{({\rm sec})}(E_{\gamma})=\left(\frac{Z}{A}\right)
I_{V}[A,\sigma^{\rm eff}_{P_{ci}N}]\left<\sigma_{{\gamma}p \to P^+_{ci}}(E_{\gamma})\right>_A
Br[P^+_{ci} \to {J/\psi}p]
\end{equation}
and the quantities $I_{V}[A,\sigma^{\rm eff}_{P_{ci}N}]$,
$\left<\sigma_{{\gamma}p \to P^+_{ci}}(E_{\gamma})\right>_A$ are defined, respectively, by Eqs. (5), (24)
of Ref. [1].
Here, $\sigma^{\rm eff}_{P_{ci}N}$ is the $P^+_{ci}$--nucleon effective absorption cross section.
This cross section can be represented as follows [1]:
%formula(28)
\begin{equation}
\sigma^{\rm eff}_{P_{ci}N}=\sigma_{P_{ci}N}+\sigma_{\rm dec}^{ci}, \,\,\,i=1, 2, 3;
\end{equation}
where $\sigma_{\rm dec}^{ci}$ is the additional to the inelastic cross section $\sigma_{P_{ci}N}$
effective $P^+_{ci}$ absorption cross section associated with their decays in the nucleus.
Using $\Gamma_{c1}=9.8$ MeV, $\Gamma_{c2}=20.6$ MeV and $\Gamma_{c3}=6.4$ MeV, we obtain
in the low-density approximation (19) that $\sigma_{\rm dec}^{c1}=2.83$ mb, $\sigma_{\rm dec}^{c2}=5.75$ mb
and $\sigma_{\rm dec}^{c3}=1.78$ mb for target nuclei considered in the case of
the free $P^+_{c1}$, $P^+_{c2}$ and $P^+_{c3}$ states production on a target proton at rest
by incident photons with resonant energies $E_{\gamma}^{\rm R1}=9.44$ GeV,
$E_{\gamma}^{\rm R2}=10.04$ GeV and $E_{\gamma}^{\rm R3}=10.12$ GeV.
For numerical simplicity, we will employ in our calculations of the cross
sections of (26), (27) for the quantity $\sigma^{\rm eff}_{P_{ci}N}$ its average value
$\sum_{i=1}^3\sigma^{\rm eff}_{P_{ci}N}$/3, which, according to Eq. (28) and with $\sigma_{P_{ci}N}=33.5$ mb,
is approximately equal to 37 mb.
%%%%%%%%%%%%%%%%%%%%%%%%%%%%%%%%%%%%%%%%%%%%%%%%%%%%%%%%%%%
\begin{figure}[htb]
\begin{center}
\includegraphics[width=16.0cm]{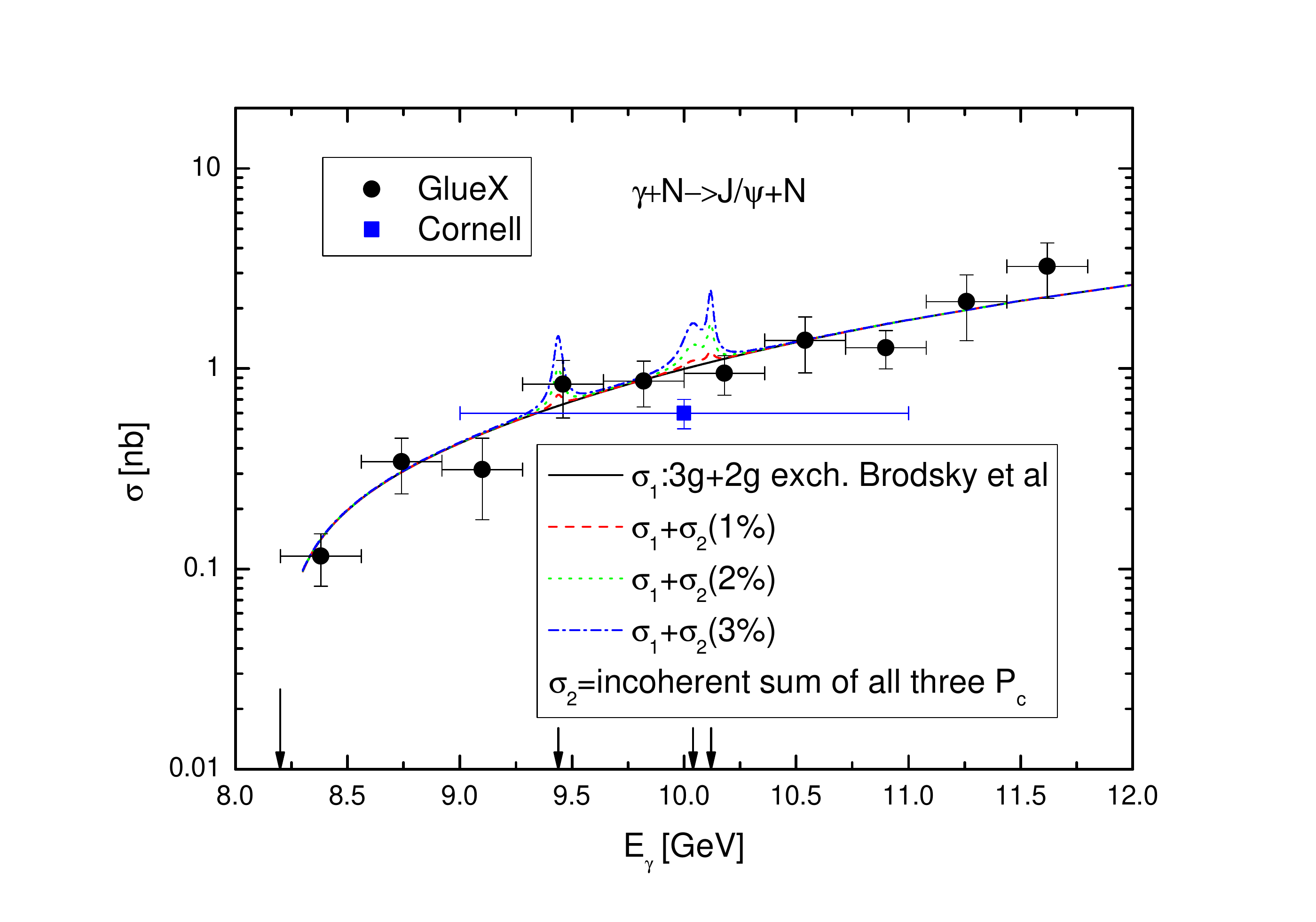}
\vspace*{-2mm} \caption{(color online) The non-resonant total cross section for the reaction
${\gamma}N \to {J/\psi}N$ (solid curve) and the incoherent sum of it and the total cross section for the
resonant $J/\psi$ production in the processes ${\gamma}p \to P^+_c(4312) \to {J/\psi}p$,
${\gamma}p \to P^+_c(4440) \to {J/\psi}p$ and ${\gamma}p \to P^+_c(4457) \to {J/\psi}p$,
calculated assuming that the resonances $P^+_c(4312)$, $P^+_c(4440)$ and $P^+_c(4457)$
with the spin-parity quantum numbers $J^P=(1/2)^-$, $J^P=(1/2)^-$ and $J^P=(3/2)^-$
decay to ${J/\psi}p$ with the lower allowed relative orbital angular momentum $L=0$
with all three branching fractions
$Br[P^+_{ci} \to {J/\psi}p]=1$, 2 and 3\% (respectively, dashed, dotted and dotted-dashed curves),
as functions of photon energy. The left and three right arrows indicate, correspondingly, the threshold energy
$E^{\rm thr}_{\gamma}=8.2$ GeV for the reaction ${\gamma}N \to {J/\psi}N$ proceeding on a free target
nucleon being at rest and the resonant energies $E^{\rm R1}_{\gamma}=9.44$ GeV,
$E^{\rm R2}_{\gamma}=10.04$ GeV and $E^{\rm R3}_{\gamma}=10.12$ GeV.
The GlueX experimental data are from Ref. [4]. The Cornell data point is from Ref. [8].}
\label{void}
\end{center}
\end{figure}
%%%%%%%%%%%%%%%%%%%%%%%%%%%%%%%%%%%%%%%%%%%%%%%%%%%%%%%%%%%%%%%%%%%%%%%%%%%%%%%%%%%%%%%%%%%%%
%%%%%%%%%%%%%%%%%%%%%%%%%%%%%%%%%%%%%%%%%%%%%%%%%%%%%%%%%%%
\begin{figure}[!h]
\begin{center}
\includegraphics[width=16.0cm]{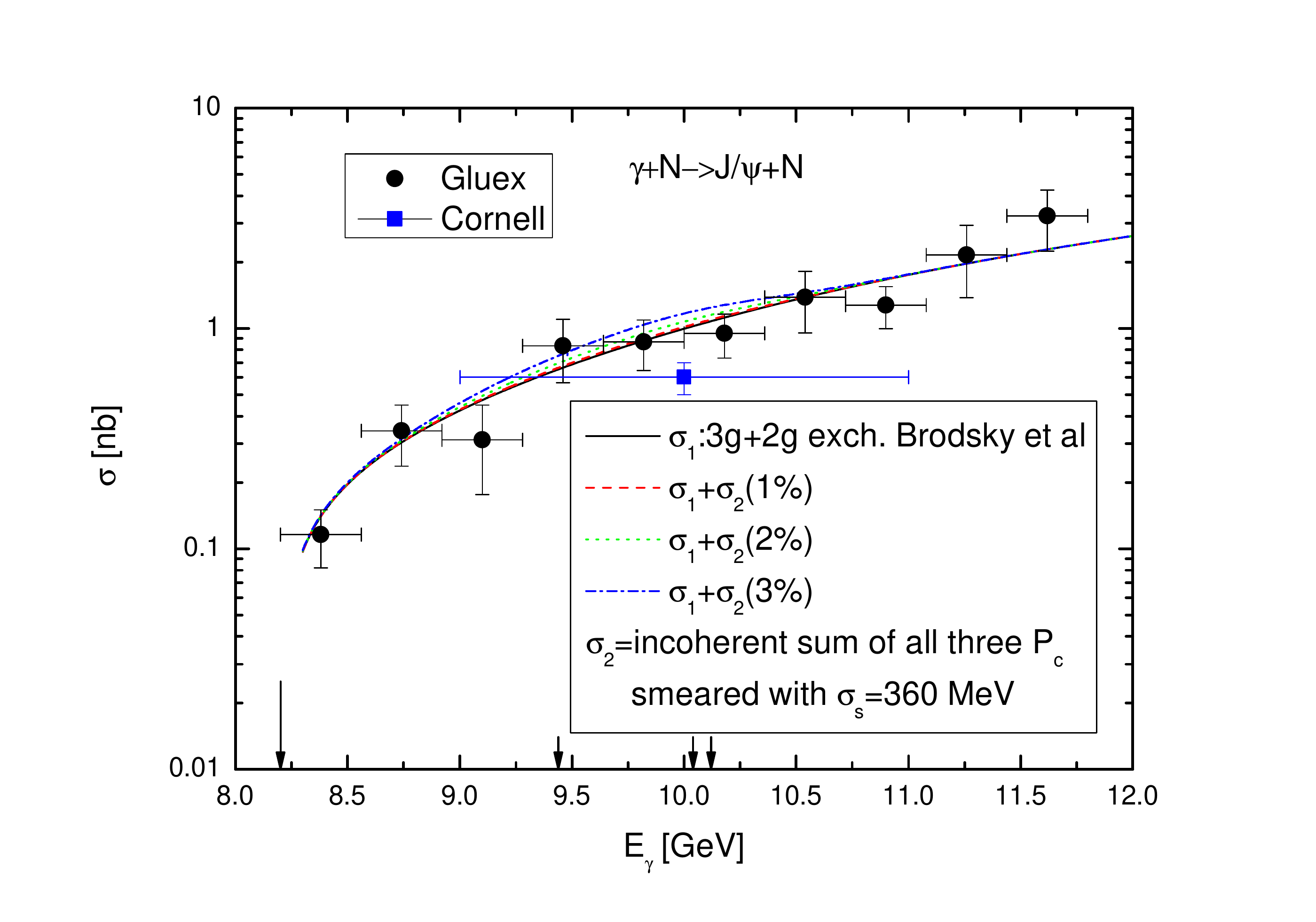}
\vspace*{-2mm} \caption{(color online) The same as in figure 2, but the contributions from the resonant channels
${\gamma}p \to P^+_c(4312) \to {J/\psi}p$, ${\gamma}p \to P^+_c(4440) \to {J/\psi}p$ and
${\gamma}p \to P^+_c(4457) \to {J/\psi}p$ are smeared by convoluting their with a Gaussian distribution
as described in the text.}
\label{void}
\end{center}
\end{figure}
%%%%%%%%%%%%%%%%%%%%%%%%%%%%%%%%%%%%%%%%%%%%%%%%%%%%%%%%%%%
%%%%%%%%%%%%%%%%%%%%%%%%%%%%%%%%%%%%%%%%%%%%%%%%%%%%%%%%%%%
\begin{figure}[!h]
\begin{center}
\includegraphics[width=16.0cm]{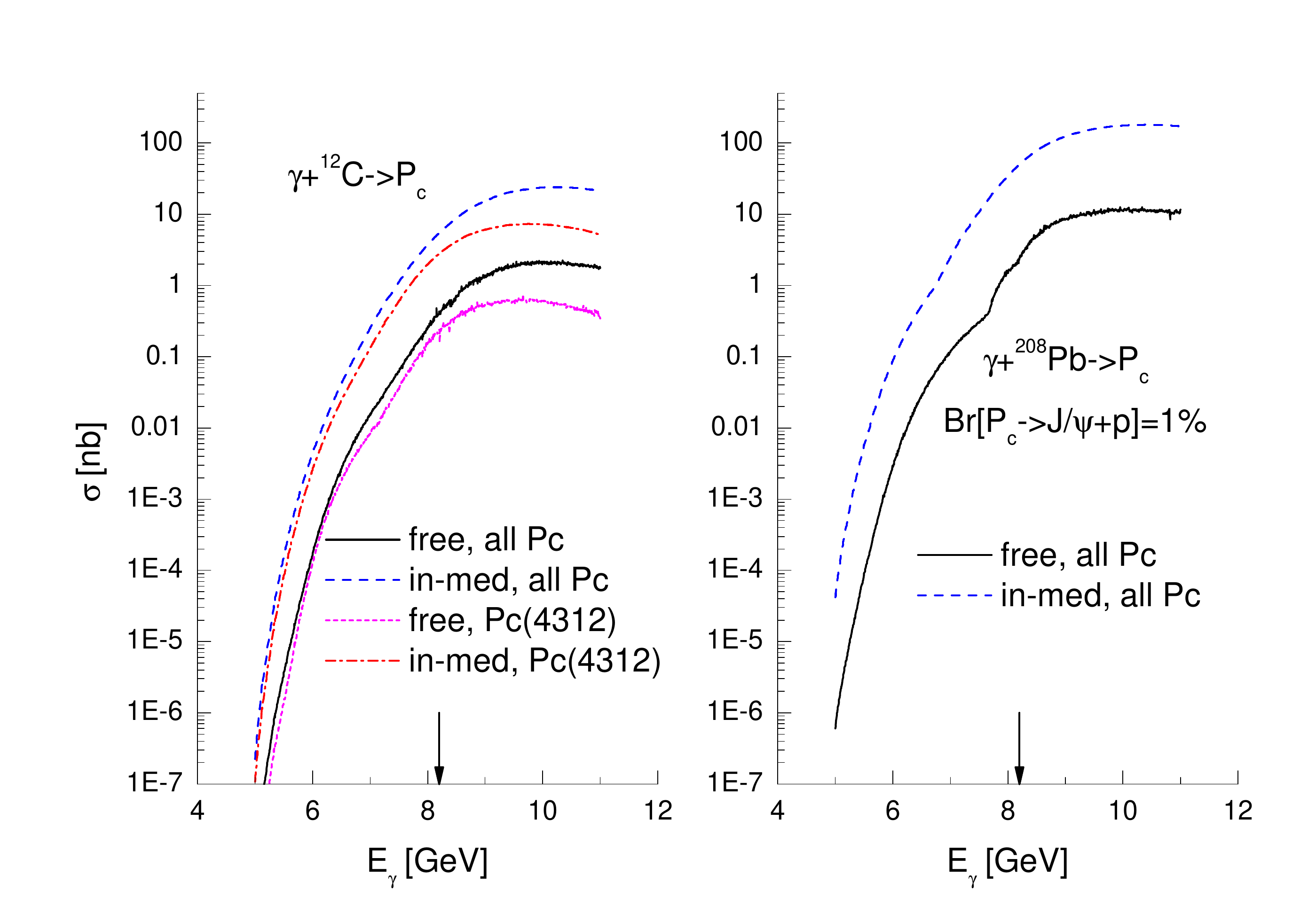}
\vspace*{-2mm} \caption{(color online) Excitation functions for resonant production of
$P^+_c(4312)$, $P^+_c(4440)$ and $P^+_c(4457)$ states
off $^{12}$C and $^{208}$Pb from the processes ${\gamma}p \to P^+_c(4312)$,
${\gamma}p \to P^+_c(4440)$ and ${\gamma}p \to P^+_c(4457)$
going on an off-shell target protons, calculated
for $Br[P^+_{ci} \to {J/\psi}p]=1$\% for all $i$ adopting free
(solid curves) and in-medium (dashed curves) $P^+_{ci}$ spectral functions.
$^{12}$C case: the same as above, but only for the process ${\gamma}p \to P^+_c(4312)$,
employing free (short-dashed) and in-medium (dotted-dashed) $P^+_c(4312)$ spectral functions.
The arrows indicate the threshold
energy for direct $J/\psi$ photoproduction on a free target nucleon being at rest.}
\label{void}
\end{center}
\end{figure}
%%%%%%%%%%%%%%%%%%%%%%%%%%%%%%%%%%%%%%%%%%%%%%%%%%%%%%%%%%%%%%%%%%%%%%%%%%%%%%%%%%%%%%%%%%%%%

\section*{3. Results and discussion}

  The free elementary non-resonant $J/\psi$ production cross section (6) (solid curve) and the combined
(non-resonant plus resonant (25)) total cross sections (dashed, dotted and dotted-dashed curves)
in comparison with the available low-energy experimental data are presented in Fig. 2. From this
figure, one can see that the $P^+_c(4312)$ state appears as clear narrow independent peak at
$E_{\gamma}=$9.44 GeV in the combined cross section, while the $P^+_c(4440)$ and $P^+_c(4457)$
resonances exhibit itself here as such two distinct narrow overlapping peaks at $E_{\gamma}=$10.04
and 10.12 GeV
\footnote{$^)$With the distance between their centroids ${\Delta}E_{\gamma}=$80 MeV.}$^)$
,
if $Br[P^+_{ci} \to {J/\psi}p]=3$\% ($i=$1, 2, 3). The strengths of these three peaks, obtained for
$Br[P^+_{ci} \to {J/\psi}p]$ $\sim$ 1--2\%, decrease essentially compared to the above case and have
a peak values, which are compatible with the respective GlueX data points. To see experimentally
such structures in the combined total cross section of the reaction ${\gamma}p \to {J/\psi}p$
one needs to have a substantially finer energy binning than that of 360 MeV adopted
in the GlueX experiment [4]. Thus, the c.m. energy ranges
$M_{ci}-{\Gamma_{ci}}/2 < \sqrt{s} < M_{ci}+{\Gamma_{ci}}/2$ ($i=$1, 2, 3) correspond to laboratory
photon energy regions of 9.416 GeV $< E_{\gamma} <$ 9.461 GeV, 9.989 GeV $< E_{\gamma} <$ 10.086 GeV
and 10.103 GeV $< E_{\gamma} <$ 10.133 GeV, i.e. ${\Delta}E_{\gamma}=$45, 97 and 30 MeV, for
$P^+_c(4312)$, $P^+_c(4440)$ and $P^+_c(4457)$, respectively. This means that to resolve the three
peaks in Fig. 2 the photon energy resolution and the energy bin size much less than at least 30 MeV
are required
\footnote{$^)$At GlueX, the $E_{\gamma}$ resolution amounts presently to 20 MeV for a 10 GeV photon [4].
Over time, it is  expected to be around 6 MeV [11]. This translates into an uncertainty
in c.m. energy of 1 MeV for a 10 GeV photon [11]. With such resolution (and bin size) will be
possible to perform a detailed scan of the $J/\psi$ total photoproduction cross section on a
proton target in the near-threshold energy region around energies $E_{\gamma}=$9.44, 10.04 and 10.12 GeV
to obtain a definite result for or against the existence of the genuine $P^+_c(4312)$, $P^+_c(4440)$
and $P^+_c(4457)$ pentaquark states and to clarify their nature and decay probabilities (cf. [26]).}$^)$
.
Otherwise, to compare correctly theoretical cross sections with the data of Fig. 2 it is necessary
to smear the contributions from the resonant channels
${\gamma}p \to P^+_c(4312) \to {J/\psi}p$, ${\gamma}p \to P^+_c(4440) \to {J/\psi}p$ and
${\gamma}p \to P^+_c(4457) \to {J/\psi}p$ by convoluting their with a Gaussian distribution
%FORMULA (29)
\begin{equation}
 G(x)=\frac{\exp{(-x^2/2{\sigma_s^2})}}{\sqrt{2{\pi}}{\sigma_s}}
\end{equation}
according to the expression [25]
%FORMULA (30)
\begin{equation}
 \left < \sigma_{{\gamma}p \to P^+_{ci}\to {J/\psi}p}(E_{\gamma},\Gamma_{ci})\right>_s
 =\int\limits_{-\infty}^{+\infty}
 \sigma_{{\gamma}p \to P^+_{ci}\to {J/\psi}p}(\sqrt{s(E_{\gamma}=y)},\Gamma_{ci})G(E_{\gamma}-y)dy,
 \,\,\,i=1, 2, 3;
\end{equation}
where ${\sigma_s}$ is the parameter of smearing. Since in the GlueX experiment the $E_{\gamma}$ bin size
is much larger than the $E_{\gamma}$ resolution, we use in smearing (30) for this parameter the value of
the bine size, i.e. ${\sigma_s}=$360 MeV. The results of such smearing of the resonant contributions in
$J/\psi$ production on proton target are given in Fig. 3.
It is seen that the peak structures of Fig. 2, corresponding to the
$P^+_c(4312)$, $P^+_c(4440)$ and $P^+_c(4457)$ states, disappear now.
The resulting combined total cross sections lie inside the experimental errors, which do not allow
to distinguish between adopted three realistic options for branching fractions
$Br[P^+_{ci} \to {J/\psi}p]$, and their shapes agree well with GlueX measurements.
It is worth noting that three-peak structure of Fig. 2, corresponding to the case of no smearing,
is also seen, as showed our calculations, when the smearing parameter ${\sigma_s}=20$ MeV is used
in Eq. (30). Now, the strengths of the peaks decrease slightly and become slightly broader compared
to those of Fig. 2. This indicates that indeed to observe the signals from the pentaquark states $P^+_{ci}$
one should choose the energy bin size $\sim$ a few tens MeV (and determine the photon energy with precision
higher than this size).

    Figure 4 shows the energy dependences of the total $P^+_c(4312)$, $P^+_c(4440)$
and $P^+_c(4457)$ production cross section in ${\gamma}$C and $\gamma$Pb reactions as well as of
the total $P^+_c(4312)$ creation in ${\gamma}$C collisions.
They are calculated on the basis of Eqs.~(26), (27)
\footnote{$^)$By assuming that in Eq. (27) $Br[P^+_{ci} \to {J/\psi}p]=1$ for all $i$ considered.}$^)$
in the scenarios with free and in-medium $P^+_c(4312)$, $P^+_c(4440)$ and $P^+_c(4457)$ spectral
functions for branching ratios $Br[P^+_{ci} \to {J/\psi}p]=1$\%. It is seen that the
resonance formation is smeared out by Fermi motion of intranuclear protons.
It is a considerably enhanced for the in-medium case at all photon energies of interest.
One can also see that at subthreshold incident energies ($E_{\gamma} < 8.2$ GeV) the main
contribution to the $J/\psi$ production on nuclei will come from the intermediate $P^+_c(4312)$ state,
while at above threshold beam energies ($E_{\gamma} > 8.2$ GeV) the resonances $P^+_c(4440)$ and
$P^+_c(4457)$ will dominate in this production.
%%%%%%%%%%%%%%%%%%%%%%%%%%%%%%%%%%%%%%%%%%%%%%%%%%%%%%%%%%%
\begin{figure}[!h]
\begin{center}
\includegraphics[width=16.0cm]{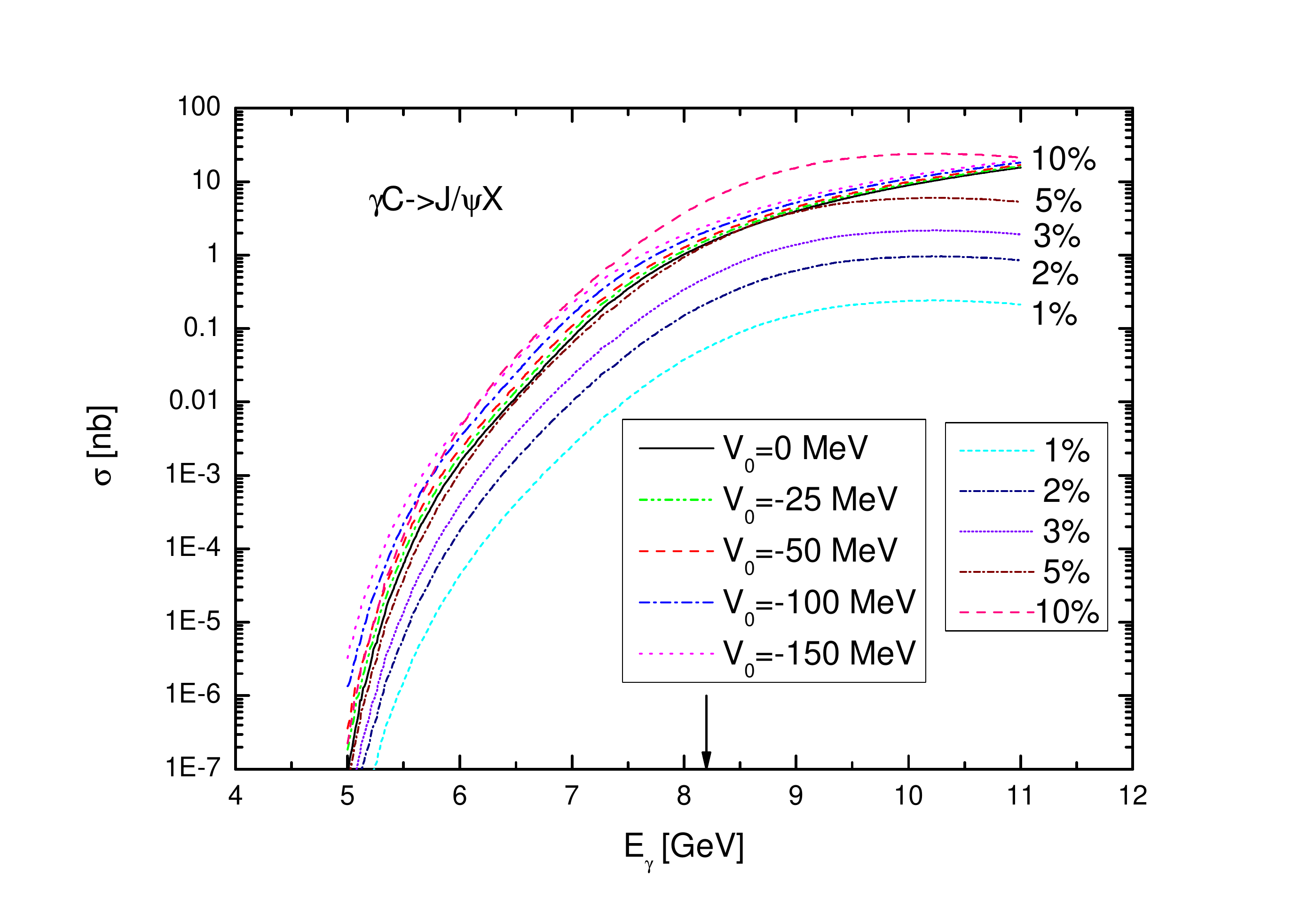}
\vspace*{-2mm} \caption{(color online) Excitation functions for the non-resonant and resonant production
of $J/\psi$ mesons off $^{12}$C from direct ${\gamma}N \to {J/\psi}N$ and resonant
${\gamma}p \to P^+_c(4312) \to {J/\psi}p$, ${\gamma}p \to P^+_c(4440) \to {J/\psi}p$ and
${\gamma}p \to P^+_c(4457) \to {J/\psi}p$
reactions going on an off-shell target nucleons. The curves,
corresponding to the non-resonant production of $J/\psi$ mesons, are calculations with an in-medium
$J/\psi$ mass shift depicted in the inset. The curves, belonging to their resonant production, are
calculations for three branching ratios $Br[P^+_{ci} \to {J/\psi}p]=1$, 2, 3, 5 and 10\% adopting
in-medium $P^+_{ci}$ spectral functions. The arrow indicates the threshold
energy for direct $J/\psi$ photoproduction on a free target nucleon being at rest.}
\label{void}
\end{center}
\end{figure}
%%%%%%%%%%%%%%%%%%%%%%%%%%%%%%%%%%%%%%%%%%%%%%%%%%%%%%%%%%%
%%%%%%%%%%%%%%%%%%%%%%%%%%%%%%%%%%%%%%%%%%%%%%%%%%%%%%%%%%%
\begin{figure}[!h]
\begin{center}
\includegraphics[width=16.0cm]{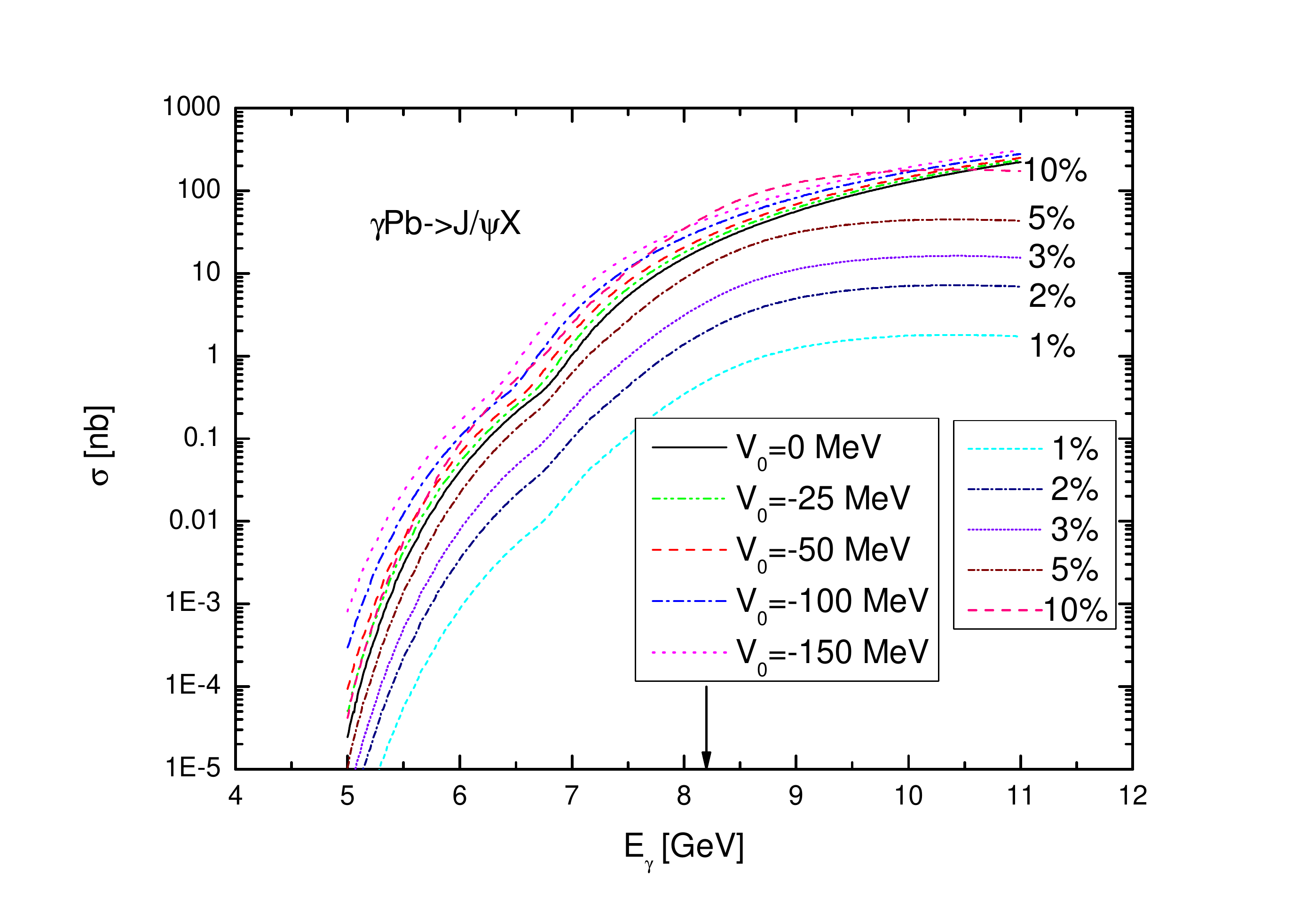}
\vspace*{-2mm} \caption{(color online) The same as in figure 5, but for the $^{208}$Pb target nucleus.}
\label{void}
\end{center}
\end{figure}
%%%%%%%%%%%%%%%%%%%%%%%%%%%%%%%%%%%%%%%%%%%%%%%%%%%%%%%%%%%

    Excitation functions for non-resonant production of $J/\psi$ mesons as well as for
their resonant production
via $P^+_c(4312)$, $P^+_c(4440)$ and $P^+_c(4457)$ resonances formation and decay in
${\gamma}$C and $\gamma$Pb reactions are given in Figs.~(5) and (6), respectively.
The former ones are calculated using Eq.~(4)
for five adopted options for the $J/\psi$ in-medium mass shift, whereas the latter ones
are determined using Eqs.~(26), (27) in the in-medium $P^+_c(4312)$, $P^+_c(4440)$ and $P^+_c(4457)$
spectral functions scenario and assuming that all three branching ratios
$Br[P^+_{ci} \to {J/\psi}p]=1$, 2, 3, 5 and 10\%.
It can be seen that in the subthreshold energy region ($E_{\gamma}$ $\sim$ 5--8 GeV) the influence of the
$J/\psi$ meson mass shift on its non-resonant yield is significant. Moreover, this influence is not masked by the
resonantly produced $J/\psi$ mesons due to the smallness here of their production cross sections compared to the
non-resonant ones, if all $Br[P^+_{ci} \to {J/\psi}p]$ are less than 5\%.
In this case, the differences between
the options $V_0=0$ MeV, $V_0=-50$ MeV, $V_0=-100$ MeV and $V_0=-150$ MeV for $J/\psi$ in-medium mass shift
can be experimentally distinguishable for both considered target nuclei
\footnote{$^)$The small mass shifts ($V_0$ $\sim$ -25 MeV) will probably be experimentally inaccessible.}$^)$
.
Thus, for incident photon energy
of 5 GeV the $J/\psi$ non-resonant yield is enhanced at mass shift $V_0=-50$ MeV by about a factor of 4.0
as compared to that obtained without this shift. When going from $V_0=-50$ MeV to $V_0=-100$ MeV and from
$V_0=-100$ MeV to $V_0=-150$ MeV the enhancement factors are about 3.5 and 2.5, respectively. At initial
beam energy of 8 GeV these enhancement factors are smaller and are about 1.3, 1.3 and 1.25. However,
the $J/\psi$ production cross sections at energy of 5 GeV are less than those at beam energy of 8 GeV by
several orders of magnitude. This makes the excitation function measurements of $J/\psi$ production at
far subthreshold photon energies (at $E_{\gamma}$ $\sim$ 5--6 GeV) a real challenge due to low cross sections
here. But at beam energies not far below the threshold (at $E_{\gamma}$ $\sim$ 7--8 GeV), the charmonium
non-resonant production cross sections have a measurable strength $\sim$ 0.1--40 nb. Therefore, the
measurements of $J/\psi$ excitation functions in ${\gamma}A$ reactions in the energy region not far below
the threshold with the aim of distinguishing at least between zero, weak ($V_0$ $\sim$ -50 MeV), relatively
weak ($V_0$ $\sim$ -100 MeV) and strong ($V_0$ $\sim$ -150 MeV) $J/\psi$ mass shifts in nuclear matter
look promising, if three branching ratios $Br[P^+_{ci} \to {J/\psi}p]$ are less than 5\%.
At above threshold photon energies of 8.2--11.0 GeV the influence of the $J/\psi$ meson mass shift on its
non-resonant yield is insignificant. Therefore, it cannot mask the contribution from the resonantly
produced $J/\psi$ mesons.
Here, the non-resonant $J/\psi$ yield (for considered options for the mass shift)
and that from the production and decay of the intermediate $P^+_c(4312)$, $P^+_c(4440)$ and
$P^+_c(4457)$ resonances are comparable for all $Br[P^+_{ci} \to {J/\psi}p]$ $\sim$ 5--10\%.
If all $Br[P^+_{ci} \to {J/\psi}p]$ are less than 5\%, then the former yield is substantially larger,
as at subthreshold energies, than the
corresponding resonant one. This means that the combined (non-resonant plus resonant) total cross
sections for $J/\psi$ production on target nuclei of interest will be enhanced compared to non-resonant
ones above threshold for all $Br[P^+_{ci} \to {J/\psi}p]$ $\sim$ 5--10\% and will not be practically
influenced here by the $J/\psi$ in-medium mass shift.
Thus, if $Br[P^+_{ci} \to {J/\psi}p]$ $\sim$ 5\% and more, the presence of three
$P^+_{ci}$ resonances in $J/\psi$ photoproduction, on the one hand, leads to additional
enhancements in the behavior of the total $J/\psi$ production cross section on nuclei above threshold,
which could be studied at JLab to get further evidence for their existence.
On the other hand, such presence can mask the modification of the mass of the non-resonantly
photoproduced $J/\psi$ mesons in nuclear matter and, therefore,
makes the determination of this modification from the excitation function measurements at
subthreshold incident energies ($E_{\gamma}$ $\sim$ 7--8 GeV) difficult.

Therefore, taking into account the above considerations, one can conclude that the $J/\psi$ excitation function
measurements at incident photon energies below the production threshold on the free target nucleon will allow
to get a deeper insight into the possible $J/\psi$ in-medium mass shifts in the range of -50 MeV and more
only if three branching ratios $Br[P^+_{ci} \to {J/\psi}p]$ are less than 5\%
\footnote{$^)$It should be pointed out again that in Ref. [1] the role of initially claimed by the
LHCb Collaboration pentaquark resonance $P^+_c(4450)$, having the quantum numbers $J^P=(5/2)^+$,
in $J/\psi$ photoproduction on nuclei near threshold was found to be insignificant only if
$Br[P^+_{c}(4450) \to {J/\psi}p]$ $\sim$ 1\% and less.}$^)$
.
It is valuable to remind that upper limits on these ratios, found by the GlueX Collaboration [4],
are within this range.

\section*{4. Conclusions}

 In this work we have calculated the absolute excitation functions for the non-resonant and resonant
photoproduction of $J/\psi$ mesons off $^{12}$C and $^{208}$Pb target nuclei in the
near-threshold beam energy region of 5--11 GeV by considering incoherent direct (${\gamma}N \to {J/\psi}N$)
and two-step (${\gamma}p \to P^+_c(4312)$, $P^+_c(4312) \to {J/\psi}p$;
${\gamma}p \to P^+_c(4440)$, $P^+_c(4440) \to {J/\psi}p$;
${\gamma}p \to P^+_c(4457)$, $P^+_c(4457) \to {J/\psi}p$)
$J/\psi$ production processes within a nuclear spectral function approach.
It was demonstrated that the non-resonant subthreshold $J/\psi$ photoproduction on nuclei reveals
well sensitivity to the possible $J/\psi$ in-medium mass shifts in the range of -50 MeV and more,
which is not masked by the resonantly produced $J/\psi$ mesons, only if
three branching ratios $Br[P^+_c(4312) \to {J/\psi}p]$, $Br[P^+_c(4440) \to {J/\psi}p]$ and
$Br[P^+_c(4457) \to {J/\psi}p]$ are less than 5\%. It was also shown that
if these branching ratios more than 5\% then the presence of the $P^+_c(4312)$,
$P^+_c(4440)$ and $P^+_c(4457)$ resonances in $J/\psi$ photoproduction produces above threshold
additional enhancements in the behavior of the total $J/\psi$ creation cross section on nuclei,
which could be also studied in the dedicated experiment at the CEBAF facility to provide
both further evidence for their existence and valuable information on their nature.
\\

%%%%%%%%%%%%%%%%%%%%%%%%%%%%%%%%%%%%%%%%%%%%%%%%%%%%%%%%%%%%%%%%
\end{document}